\documentclass[conference]{IEEEtran}

\usepackage{cite}
\usepackage{balance}

\ifCLASSINFOpdf
\usepackage[pdftex]{graphicx}
\usepackage{epstopdf}
\usepackage{subcaption}
\usepackage{pgfplots}
\usepackage{amsmath,stmaryrd,pict2e,picture}
\else
\fi
\usepackage{amsmath,amssymb,amsthm}
\usepackage{mathrsfs}
\usepackage{algorithm, algorithmic}

\def\footnoterule{\relax%
	\kern-5pt
	\hbox to \columnwidth{\hfill\vrule width 0.75\columnwidth height 0.4pt\hfill}
	\kern4.6pt}
\makeatother

\IEEEoverridecommandlockouts

\begin{document}
	%
	\title{ Sparse Array DFT Beamformers  for Wideband Sources}
	%
	%
	%
	\author{\IEEEauthorblockN{Syed~A.~Hamza and
			Moeness~G.~Amin}
		\IEEEauthorblockA{ Center for Advanced Communications, 
			Villanova University, Villanova, PA 19085, USA\\
			Emails: \{shamza, moeness.amin\}@villanova.edu}
\thanks{This work  is supported by NSF award \#AST- 1547420.}}
\maketitle
\begin{abstract}
Sparse arrays  are popular for  performance optimization while keeping  the hardware and computational costs down. In this paper, we consider   sparse arrays design method  for wideband  source  operating in a wideband jamming environment. Maximizing the  signal-to-interference plus  noise ratio (MaxSINR) is adopted as an optimization objective for wideband beamforming. Sparse array design problem is   formulated in the DFT domain  to process the source as parallel narrowband sources. The  problem is formulated  as quadratically constraint quadratic program (QCQP) alongside the  weighted mixed $l_{1-\infty}$-norm squared penalization of the beamformer weight vector. The semidefinite relaxation (SDR) of QCQP  promotes sparse solutions by iteratively re-weighting beamformer based on  previous iteration. It is shown that the DFT approach reduces the computational cost considerably as compared to the delay line approach, while efficiently utilizing the degrees of freedom to harness the maximum output SINR offered by the given array aperture.
\end{abstract}
\IEEEpeerreviewmaketitle
\section{Introduction}
Sparse array design strives to optimally deploy  sensors, essentially   achieving desirable beamforming characteristics, lowering  the system hardware costs and  reducing the computational  complexity \cite{1139138}.  Designing sparse arrays for wideband signal models can potentially offer considerable savings  in hardware and data storage needed to process data jointly in the spatial and temporal domains. Sparse array optimum configuration design is  primarily guided by environment-independent design objectives, such as desirable beampattern characteristics and enhancing   source identifiability \cite{5739227, AMIN20171, 5456168}. Recent development in sparse array design has  considered environment-dependent type of objectives. This has been made possible by the emerging fast and cost-effective antenna switching technologies.  In this case, optimum sparse array is the one that achieves and maintains performance optimality under various sensing conditions using a given and limited number of  sensors within the available aperture. One of key optimality criteria is maximizing the  signal-to-interference plus  noise ratio  (SINR)   which has been quite successful in yielding array configurations resulting in enhanced target parameter  estimation accuracy \cite{663844798989898, 6774934, 7944303, 8061016}.   
 
In this paper, we consider environment-dependent  MaxSINR sparse arrays design for wideband sources   in presence of  wideband jammers. This is in contrast with environment-independent wideband beamforming for sparse arrays which has been investigated for frequency independent beampattern synthesis and sidelobe level control  \cite{1071, doi:10.1121/1.412215, 93229}.  In essence, we  adopt a Capon based methodology which is data dependent beamforming aiming at  enhancing the   desired signal power and reducing undesired signal components at the array output when operating    in an interference active environment \cite{1206680}. 

Wideband sources are commonly encountered in multitude of applications in array  processing \cite{1359140, 2f1ce6c9e62d4b60acddd36683f19924}. Wideband beamforming is executed either as a delay line filtering or DFT implementation \cite{1450747, 1164219, 1143832, 1165142}.  In this paper, we focus on the latter where the data at each sensor is buffered and transformed to the frequency domain by $L$-point DFT. In this case, the optimal beamformer seeks to maximize SINR in each frequency bin individually. The underlying problem is then cast as finding the optimum array configuration across all frequency bins which maximizes the SINR at the array output.    
To underscore sparsity in the spatial domain, we pose  the   problem as optimally selecting $P$ antennas out of $N$ possible equally spaced locations. Then, the optimum Capon beamformer is the one that achieves the design objective, considering all possible sparse array configurations that stem from different arrangements of the available antennas. 

It has been shown that for uniform linear arrays (ULAs), Capon beamformer maximizes the principal eigenvalue of the product of the received data correlation and  desired correlation matrix \cite{1223538}.  The same principle applies to the underlying problem where such maximization involves two sets of variables pertaining to sensor placements and multiple frequencies.    It is noted, however, that  principal eigenvalue maximization    is a  combinatorial optimization and is, therefore,  NP hard. In order to avoid the computational burden of singular value decomposition (SVD) for each possible array configuration, we solve the  problem by convex approximation.

 The design problem at hand is posed as QCQP with weighted mixed $l_{1-\infty}$-norm  penalization. This formulation promotes group sparsity to ensure that  $P$ antenna sensors are selected in the beamforming weight vectors corresponding to $L$ DFT bins. We opt to use $l_{1-\infty}$-norm squared penalization, as a regularization term, which naturally leads to the  semidefinite relaxation (SDR). The SDR is the convex relaxation of QCQP and, therefore, can be solved in polynomial time.   It is shown that the solution of the underlying QCQP is subsequently  given by the principal eigenvector of the SDR solution matrix.    In order to promote  rank one SDR solutions iteratively, and recover sparse solutions effectively,  we  adopt a modified eigenvector based   scheme to update the regularization weighting matrix. The proposed approach builds on the fact that the weighted $l_1$-norm convex relaxation has been  exploited for antenna selection problems    for  beampattern synthesis, whereas the weighted $l_{1-\infty}$-norm squared relaxation   has been shown to effectively reduce the required number of antennas in multicast transmit beamforming \cite{6663667, article1, 6477161}. We demonstrate the offerings of the proposed sparse array design approach by comparing its performance with those    sparse arrays  developed by  existing design methods. \par

The rest of the paper is organized as follows:  In the next section, we  state the problem formulation for  maximizing the  output SINR under wideband source signal model by explaining the DFT signal model in detail. Section \ref{Optimum sparse array design} deals with the optimum sparse array design by  semidefinite relaxation  to find the optimum $P$  antenna sparse array geometry. Simulations are presented for  the above cases, and conclusion follows at the end.

\section{Problem Formulation} \label{Problem Formulation}
A  desired source and $Q$ interfering source signals impinging on a  linear array with $N$ uniformly placed antennas. The    baseband signal $\mathbf{x}(n) \in \mathbb{C}^{N}$ having bandwidth  $B_s/2$,  is sampled at the receiver at the Nyquist rate. The received signal at time instant $n$ is therefore  given by: 
\begin {equation} \label{a}
\mathbf{x}(n)=    \mathbf{s}(n)  + \sum_{k=1}^{Q}  \mathbf{i}_k(n) + \mathbf{v}(n),  
\end {equation}
where $\mathbf{s}(n) $ is the contribution from the desired signal located at $\theta_s$, $\mathbf{i}_k(n) $  are  the interfering signal vectors  corresponding to the  respective directions  of arrival, $\theta_i$ and $\mathbf{v}(n)$ is the spatially uncorrelated sensor array output noise.\\
The received signal $\mathbf{x}(n)$ is processed in the spectral domain by taking an $L$ point DFT for the data received by $k$th sensor ${x_k}(n)$, 
\begin{equation} \label{l}
{X}_k(l) = \sum_{p=1}^{L}x_k(n-p)(e^{-j\frac{2\pi}{L}})^{lp}, \, \, \, \, \, \, l=0,1, . . . \, L-1.
\end{equation}
Define a vector ${\mathbf{X}}_l \in C^N$, containing the  $l$th DFT bin data corresponding to each sensor, 
\begin {equation} \label{m}
{\mathbf{X}}_l= [{X}_1(l), {X}_2(l), . . ., {X}_N(l) ]^T.
\end {equation}
The data for the $l$th data bin is then combined linearly by weight vector $\mathbf{w}_{l}$ such that,
\begin {equation}  \label{n}
{y}_l = \mathbf{w}_l^H \mathbf{X}_l,  \, \, \, \, \, \, l=0,1, . . . \, L-1.
\end {equation}
Subsequently, the overall beamformer output $y$ is generated by taking the inverse DFT of $y_l$ generated across $L$ beamformers.
 The DFT implementation seeks to maximize the output SINR for each frequency bin, yielding the following optimization problem,
  \begin {equation}  \label{o}
 \begin{aligned}
 \underset{\mathbf{w}_0, \mathbf{w}_1, ... \mathbf{w}_{L-1}}{\text{minimize}} & \quad   \sum_{l=1}^{L}\mathbf{ w}_{l}^H\mathbf{R}_{l}\mathbf{ w}_{l},\\
 \text{s.t.} & \quad    \sum_{l=1}^{L} \mathbf{ w}_l^H\mathbf{R}_{sl}\mathbf{ w}_l=1. 
 \end{aligned}
 \end {equation}
 The correlation matrix $\mathbf{R}_{l}=\mathbf{X}_l\mathbf{X}_l^H$ is the received  correlation matrix for the  $l$th processing bin. Similarly, the source  correlation matrix $\mathbf{R}_{sl}$ for the desired source impinging from direction of arrival $\theta_s$ is given by,
 \begin {equation} \label{p}
\mathbf{R}_{sl}=\mathbf{S}_l\mathbf{S}_l^H=\sigma_{sl}^2\mathbf{a}(\theta_s,l)\mathbf{a}^H(\theta_s,l) 
 \end {equation}
 Here, $\mathbf{S}_l$ is the received data vector representing the desired source in the $l$th  bin  and DOA $\theta_s, \sigma_{sl}^2$ donates the power of this source,  $\mathbf{a}(\theta_s,l)$  is  the corresponding  steering vector for the source and is defined as follows,
   \begin {equation} \label{q}
 \begin{aligned}
 \mathbf{a}(\theta_s,l)= {}&[1 \,  \,  \, e^{j \pi( \frac{\Omega_{min}+l\Delta_{\omega}}{\Omega_{max}}) cos(\theta_s)}  {}\\
  &\,  . \,   . \,  . \, \,  \,  \, \,  \,  \, e^{j \pi( \frac{\Omega_{min}+l\Delta_{\omega}}{\Omega_{max}} ) (N-1)cos(\theta_s)}]^T.
 \end{aligned}
 \end {equation}
 Equation (\ref{q}) models the steering vector for the $l$th frequency bin, where $\Omega_{min}=w_c-\frac{B_s}{2}$ is the lower edge of the passband frequency and $\Delta_{\omega}=\frac{B_s}{L}$ is the frequency resolution, $w_c$ being the carrier frequency. 
Similar to the tapped delay line,  the DFT implementation can equivalently determine the the optimum sparse array geometry for enhanced MaxSINR performance as explained in the following section.
\section{Optimum sparse array design}  \label{Optimum sparse array design}
The problem of maximizing the principal eigenvalue of the  correlation matrices associated with $P$ antenna selection is a combinatorial optimization problem. We assume  the knowledge of the  full array data correlation matrix which is realizable for fully-augmentable sparse arrays and estimated by correlation matrix completion and interpolation schemes. We  formulate the sparse array design for MaxSINR in case of wideband beamforming  as a rank relaxed  semidefinite program (SDR). 
 \subsection{Semidefinite Programming for Sparse solution}
  We assume that the antenna configuration remain the same within the observation time.  Therefore, it is required that the same $P$ antennas are selected for  each DFT bin within the coherent processing  interval.  We  optimally pick $P$ entries from the beamforming weight vector for the first DFT bin and the same  $P$ entries are to be selected for the remaining $L-1$ frequencies.  Define $\mathbf{w}_k \in \mathbb{C}^L$ to be the weights for all the $L$ DFT bins corresponding  to the $ k$th sensor.   Then, rewrite the problem formulated in (\ref{o})  as follows:
\begin{equation} \label{a2}
\begin{aligned}
 \underset{\mathbf{w}_l \mathbf{\in \mathbb{C}}^{N}}{\text{minimize}} & \quad   \sum_{l=1}^{L}\mathbf{ w}_{l}^H\mathbf{R}_{l}\mathbf{ w}_{l} + \mu(\sum_{k=1}^{N}||\mathbf{w}_k||_q),\\
\text{s.t.} & \quad    \sum_{l=1}^{L} \mathbf{ w}_l^H\mathbf{R}_{sl}\mathbf{ w}_l=1. 
\end{aligned}
\end{equation}
Here,  $||.||_q$ denotes the $q$-norm of the vector. The mixed $l_{1-q}$ norm regularization  is know to thrive  the group sparsity in the solution for $q>1$ as is required in our case. The relaxed  problem expressed in Eq. (\ref{a2}) induce the group sparsity in optimal weight vectors $\mathbf{w}_l$ without placing a hard constraint on the specific cardinality of $\mathbf{w}_l$. 
 The  problem in $(\ref{a2})$  can be  penalized instead  by the weighted $l_1$-norm function  which is a well known sparsity promoting formulation \cite{Candès2008},
\begin{equation} \label{b2}
\begin{aligned}
\underset{\mathbf{w}_l \mathbf{\in \mathbb{C}}^{N}}{\text{minimize}} & \quad   \sum_{l=1}^{L}\mathbf{ w}_{l}^H\mathbf{R}_{l}\mathbf{ w}_{l}  + \mu(\sum_{k=1}^{N}\mathbf{u}^i(k)||\mathbf{w}_k||_q),\\
\text{s.t.} & \quad    \sum_{l=1}^{L} \mathbf{ w}_l^H\mathbf{R}_{sl}\mathbf{ w}_l=1. 
\end{aligned}
\end{equation}
where, $\mathbf{u}^i(k)$  is the $k$th element of  re-weighting  vector $\mathbf{u}^i$ at  the $i$th iteration. We choose the  $\infty$-norm  for the  $q$-norm and  replace the weighted $l_1$-norm function   in $(\ref{b2})$  by the  $l_1$-norm  squared  function without effecting it's regularization property  \cite{6477161},
\begin{equation} \label{c2}
\begin{aligned}
 \underset{\mathbf{w}_l \mathbf{\in \mathbb{C}}^{N}}{\text{minimize}} & \quad   \sum_{l=1}^{L}\mathbf{ w}_{l}^H\mathbf{R}_{l}\mathbf{ w}_{l}  +   \mu(\sum_{k=1}^{N}\mathbf{u}^i(k)||\mathbf{w}_k||_\infty)^2,\\
\text{s.t.} & \quad    \sum_{l=1}^{L} \mathbf{ w}_l^H\mathbf{R}_{sl}\mathbf{ w}_l=1. 
\end{aligned}
\end{equation}
The SDR for the above problem can then be realized  by re-expressing the quadratic functions, $ \mathbf{ w}_l^H\mathbf{R}_l\mathbf{w}_l= $ Tr$(\mathbf{ w}_l^H\mathbf{R}_l\mathbf{w}_l)= $Tr$(\mathbf{R}_l\mathbf{w}_l\mathbf{ w}_l^H)=$ Tr$(\mathbf{R}_l\mathbf{W}_l)$, where Tr(.) is the trace of the matrix. This expression yields the following problem \cite{Bengtsson99optimaldownlink, 5447068, 6477161},
 \begin{equation} \label{e2}
\begin{aligned}
\underset{\mathbf{W}_l \mathbf{\in \mathbb{C}}^{N.N}, \mathbf{\tilde{W} \in \mathbb{R}}^{N.N}}{\text{minimize}} &
 \quad   \sum_{l=1}^{L}\text{Tr}(\mathbf{R}_l\mathbf{W}_l) + \mu \text{Tr}(\mathbf{U}^i\mathbf{\tilde{W}}),\\
\text{s.t.} & \quad    \sum_{l=1}^{L}\text{Tr}(\mathbf{R}_{sl}\mathbf{W}_l) \geq1,  \\
& \quad \mathbf{\tilde{W}} \geq  |\mathbf{W}_{l}|  \quad \forall \, \, l \in \,\,  0, 1, . . .  L-1.    ,\\
& \quad   \mathbf{W} \succeq 0, \, \text{Rank}(\mathbf{W})=1.
\end{aligned}
\end{equation}
Here $'\geq'$ is the element wise comparison and $'\succeq'$ represents inequality in the matrix sense, $\mathbf{W}_{l} \in \mathbb{C}^{N.N}$ is the outer product of the  $l$th beamforming weight vector, $\mathbf{W}_l=\mathbf{w}_l\mathbf{ w}_l^H$ and  $\mathbf{U}^i=\mathbf{u}^i(\mathbf{u}^i)^T$. The  rank constraint in  Eq. (\ref{e2}) is non convex and, therefore, we drop the rank constraint   resulting in the following  SDR:
 \begin{equation} \label{f2}
\begin{aligned}
\underset{\mathbf{W}_l\mathbf{\in \mathbb{C}}^{N.N}, \mathbf{\tilde{W} \in \mathbb{R}}^{N.N}}{\text{minimize}} & \quad   \sum_{l=1}^{L}\text{Tr}(\mathbf{R}_l\mathbf{W}_l) + \mu \text{Tr}(\mathbf{U}^i\mathbf{\tilde{W}}),\\
\text{s.t.} & \quad    \sum_{l=1}^{L}\text{Tr}(\mathbf{R}_{sl}\mathbf{W}_l) \geq1,  \\
& \quad \mathbf{\tilde{W}} \geq  |\mathbf{W}_{l}|  \quad \forall \,\, l \in \,\, 0, 1, . . .  L-1.    ,\\
& \quad   \mathbf{W} \succeq 0.
\end{aligned}
\end{equation}
\begin{algorithm}[t!] \label{algorithm}
	
	\caption{SDR for optimal sparse beamforming vectors.}
	
	\begin{algorithmic}[]
		
		\renewcommand{\algorithmicrequire}{\textbf{Input:}}
		
		\renewcommand{\algorithmicensure}{\textbf{Output:}}
		
		\REQUIRE Received data correlation matrix $\mathbf{R}_l$'s, $N$, $P$, $L$, look direction DOA $\theta_s$. \\
		
		\ENSURE  $L$  beamforming weight vectors.   \\
		
		\textbf{Initialization:} \\
		
		Initialize $\mu$, $\epsilon$, $\mathbf{U}$ is all ones matrix.
		
		\WHILE {(Sparsity is not invoked in $|\mathbf{\tilde{W}|}$ )}
		
		\STATE   Run the SDR of Eq. (\ref{f2}).\\ 
		\STATE   Update the regularization weighting matrix $\mathbf{U}$ according to Eq. (\ref{h2}). \\
		\ENDWHILE \\
		Binary search for desired cardinality $P$\\
		$l=\mu_{lower}$, $u=\mu_{upper}$ (Initializing lower and upper limits of sparsity parameter range)
		\WHILE {(Cardinality of $\mathbf{w}_l$ $\neq$ $P$)}
		\STATE  $\mu=[(l+u)/2]$
		\STATE   Run the  SDR of Eq. (\ref{f2}) with the last regularization weighting matrix $\mathbf{U}$ from the first while loop.\\ 
		
		\IF { (Cardinality of $\mathbf{w}_l$) $<$ $P$}
		\STATE  $u=\mu$
		\ELSE
		\STATE  $l=\mu$
		\ENDIF
		
		\ENDWHILE
		
		\STATE After achieving the desired cardinality, run SDR for reduced size correlation matrix corresponding to nonzero values of $\mathbf{\tilde{W}}$  and $\mu=0$,  yielding, $\mathbf{w}_l=\mathscr{P} \{  \mathbf{W}_l \} $. 
		
		\RETURN $\mathbf{w}_l$
		
	\end{algorithmic}
	\label{algorithm}
\end{algorithm}
 It is apparent from the problem formulation  that the DFT approach involves $L$ unknown variables of dimension $N*N$, whereas, the delay line filtering approach has the  dimensionality of the  order of  $NL*NL$  that  makes the DFT approach  computationally more  viable.
 
 As suggested in \cite{Candès2008}, the weight matrix  $\mathbf{U}^i$ is initialized unweighted, i.e., a matrix of all ones. It is iteratively updated as follows,
\begin{equation} \label{g2}
\mathbf{U}_{m,n}^{i+1}=\frac{1}{\mathbf{\tilde{W}}^i(m,n)+\epsilon}.
\end{equation}
The parameter $\epsilon$ prevents  the unwanted case of division by zero and also avoids the solution to converge to  local minima. The $m,n$th entry of $\mathbf{\tilde{W}}$ is given by $\mathbf{\tilde{W}}^i(m,n)$. However, for the underlying problem,  the solution matrices $\mathbf{W}_l$   is not exactly rank  one matrix at each iteration. Therefore, the weight matrix iteratively favors solution of higher ranks and struggles to yield desirable sparse solutions. To mitigate this problem, we  approximate the solution matrix by rank $1$ approximation as, 
\begin{equation} \label{h2}
\mathbf{U}_{m,n}^{i+1}=\frac{1}{\mathbf{Y}^i(m,n)+\epsilon}.
\end{equation}
where, $\mathbf{Y}^i=\mathbf{y}^i(\mathbf{y}^i)^T,$ for $\mathbf{y}^i=\frac{1}{L}\sum_{l=1}^{L}(\mathbf{|\mathscr{P}\{W}_l^i\}|)^2 $. The operator $\mathscr{P} \{. \}$  denotes the principal eigenvector of the input matrix. Clearly, $\mathbf{Y}^i$ is rank one matrix.  This modified reweighing approach effectively solves the  optimum sparse array  selection.   The proposed algorithm for controlling the sparsity of the optimal weight vector is summarized in  Algorithm. 1.

\begin{figure}[!t]
	\centering
	\includegraphics[height=1.95in, width=3.75in]{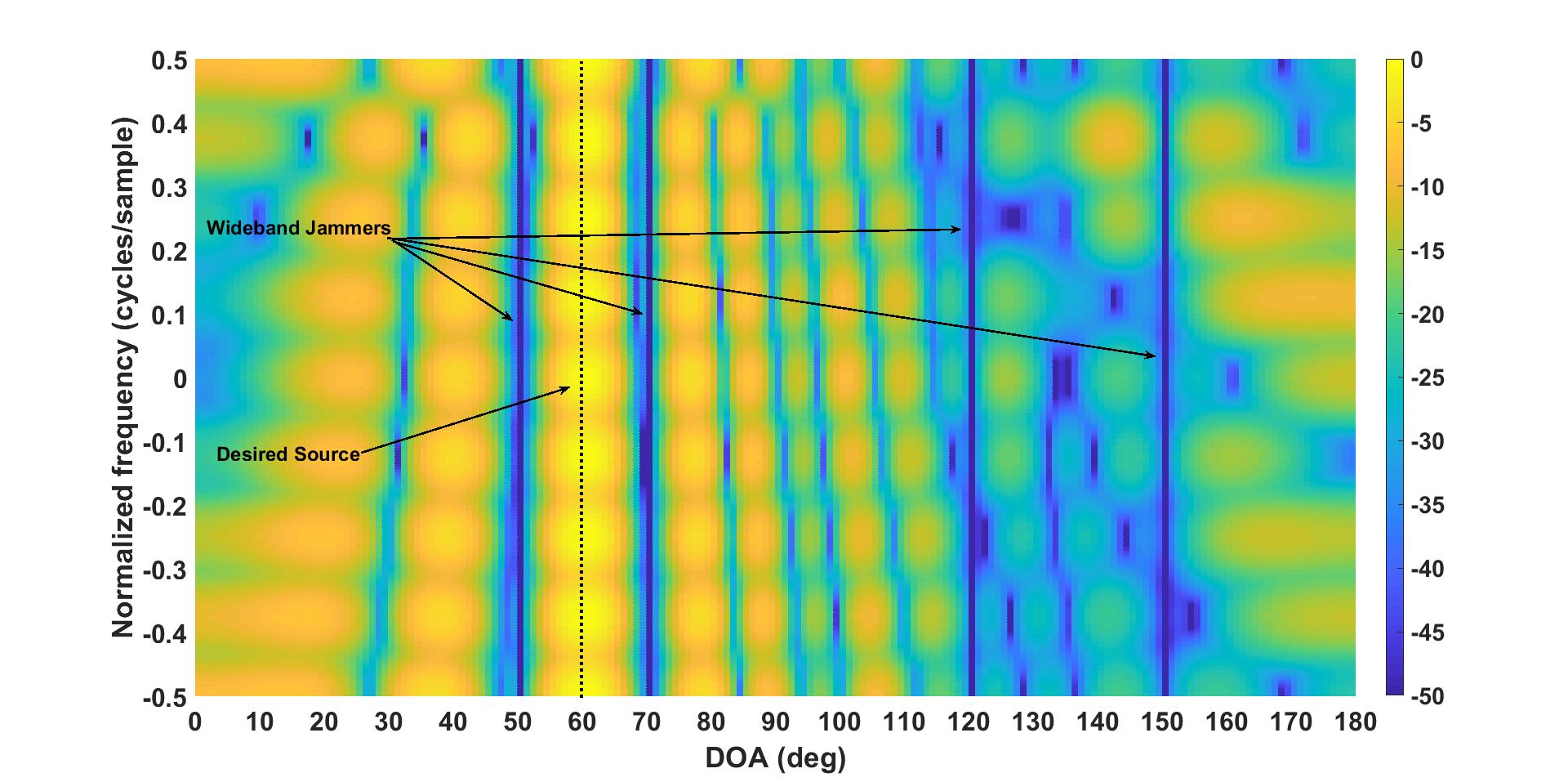}
	\caption{Frequency dependent beampattern for the optimum array recovered through enumeration.}
	\label{op_algo_2027_8}
\end{figure}

\begin{figure}[h]
	\centering
	\begin{subfigure}[b]{0.45\textwidth}
		\includegraphics[height=0.27in, width=1\textwidth]{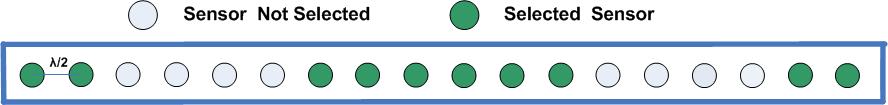}
		\caption{}
		\label{op_enumer_608}
	\end{subfigure}
	\begin{subfigure}[b]{0.45\textwidth}
		\includegraphics[height=0.17in, width=1\textwidth]{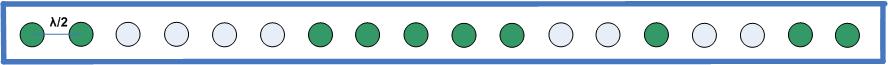}
		\caption{}
		\label{op_notap_775}
	\end{subfigure}
	\begin{subfigure}[b]{0.45\textwidth}
		\includegraphics[height=0.17in, width=1\textwidth]{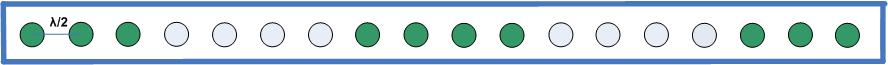}
		\caption{}
		\label{op_algo_2027}
	\end{subfigure}
	\begin{subfigure}[b]{0.45\textwidth}
		\includegraphics[height=0.17in, width=1\textwidth]{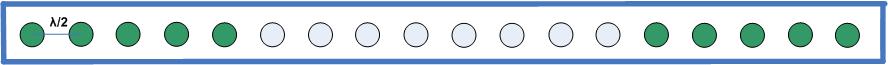}
		\caption{}
		\label{wo_enu_924}
	\end{subfigure}
	\caption{(a) Optimum $10$ antenna array for DFT implementation (SDR)  (b)  Optimum $10$ antenna array for DFT implementation  (enumeration)   (c)  Optimum $10$ antenna array for delay line  implementation (SDR)  (d)  Worst performance $10$ antenna array  for DFT implementation}
	\label{op_c_co_c}
\end{figure}
 \section{Simulations} \label{Simulations}
We demonstrate the effectiveness of the sparse array design for  MaxSINR by adopting  the DFT beamformer approach.  The  performance   of   the DFT beamformer  for MaxSINR is further compared with the   delay line filtering   to process wideband signals optimally. 

\subsection{Example 1}
We select $P=10$ sensors from $N=18$  possible equally spaced locations with  inter-element spacing of $\lambda_{min}/2$. The array data is sampled periodically at the Nyquist rate. We consider   $8$ DFT bins for DFT implementation, implying $L=8$ ($8$  filter taps associated  with each selected antenna sensor for delay line implementation). 
\begin{figure}[!t]
	\centering
	\includegraphics[height=1.95in, width=3.75in]{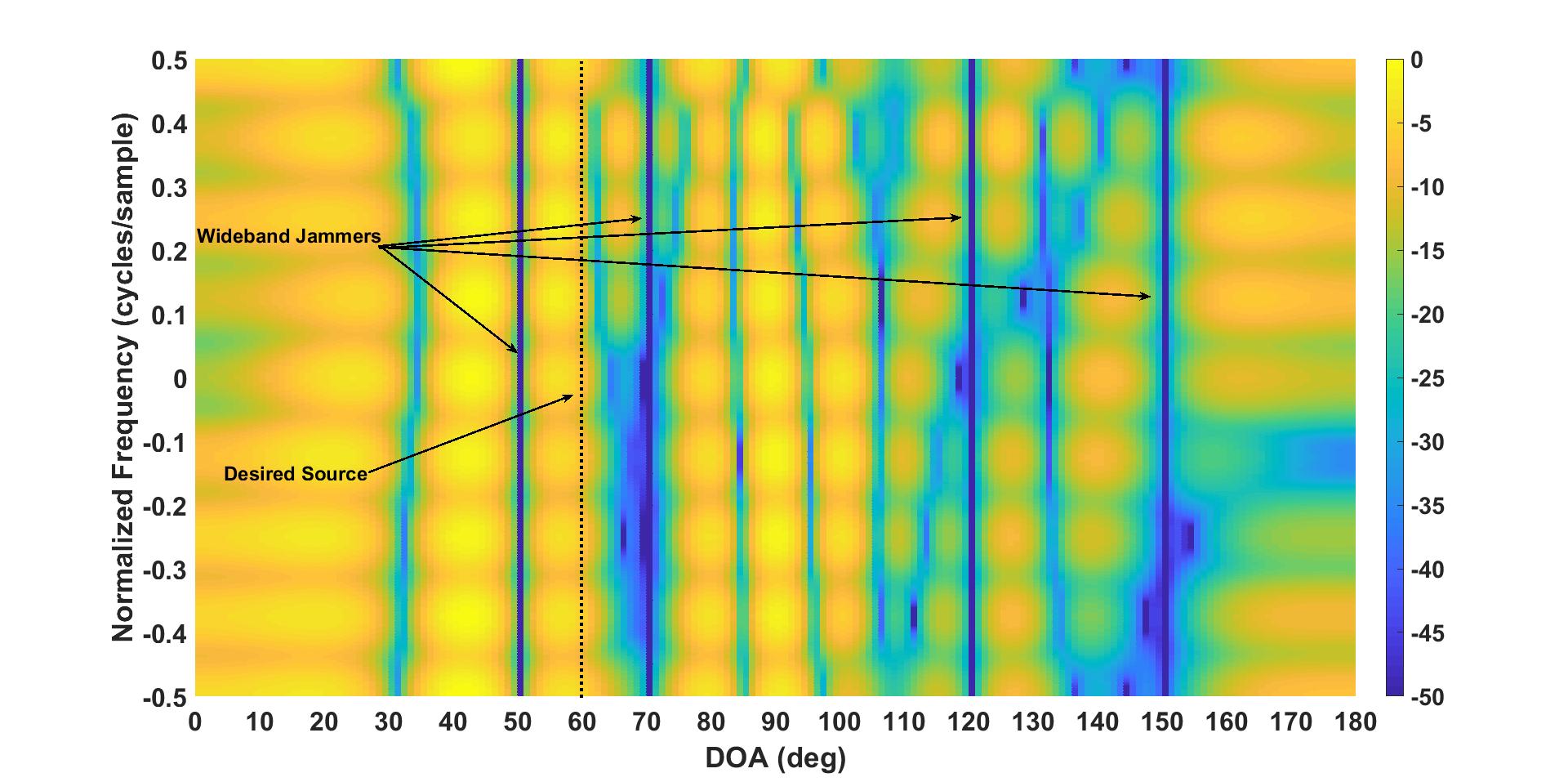}
	\caption{Frequency dependent beampattern for the worst case sparse array topology.}
	\label{uni_uni_1_8}
\end{figure}
A frequency spread   desired point source   is  impinging on a linear array from  DOA $60^0$. The PSD of the frequency spread source is uniform from -0.5  Hz to 0.5 Hz. Four strong wideband interferers  with the  uniform PSD  are operating  from $50^0$, $70^0$, $120^0$ and $150^0$.   The SNR of the desired signal is $0$ dB, and the INR of  each interfering signals is set to $30$ dB. The fractional bandwidth of the source is  such that the maximum normalized spatial frequency is $0.5$. Figure \ref{op_algo_2027_8} shows the frequency dependent beampattern  for the  optimum array configuration recovered through SDR. The beampattern depicts the maximum  gain throughout the frequency band occupied by the source of interest at $60^0$, whereas,  the interferers face an attenuation of greater than $50$ dB for all possible frequencies. Therefore, the optimum  sparse array  configuration recovered through SDR (array topology shown in the Fig. (\ref{op_enumer_608})) delivers a promising SINR performance of $9.85$ dB.  The optimum sparse array found through exhaustive search (shown in the Fig. (\ref{op_notap_775})) offers an output SINR of $9.88$ dB that is sufficiently close to the array yielded by the convex relaxation. It is  noted that   the exhaustive search involves expensive  singular value decomposition (SVD)  for $43758$ possible configurations, which has very high computational cost.  Figure (\ref{op_algo_2027}) shows the optimum sparse array achieved through SDR for the delay line  scheme for MaxSINR wideband beamforming. This array configuration offers an output SINR of $9.7$ dB and is lower than the sparse array design realized for the DFT beamformer implementation. However, the maximum possible SINR offerings in the delay line implementation is $9.94$ dB (found through exhaustive search) and is greater than the maximum SINR offered by the DFT approach.   Figure (\ref{wo_enu_924}) depicts the sparse array configuration with the worst case SINR of $1.7$ dB. This considerable performance degradation is explained from  Fig. (\ref{uni_uni_1_8}) which shows the beampattern  associated with the worst case sparse array configuration. In this case, the optimal weights  strive to alleviate the high power jammers but   struggle to place the maxima towards the source of interest, thereby losing considerably in SINR performance.  It is of interest to observe that   the worst case sparse array configuration occupies the entire available aperture yet compromises significantly on the  performance.

\begin{figure}[!t]
	\centering
	\includegraphics[height=1.85in, width=3.5in]{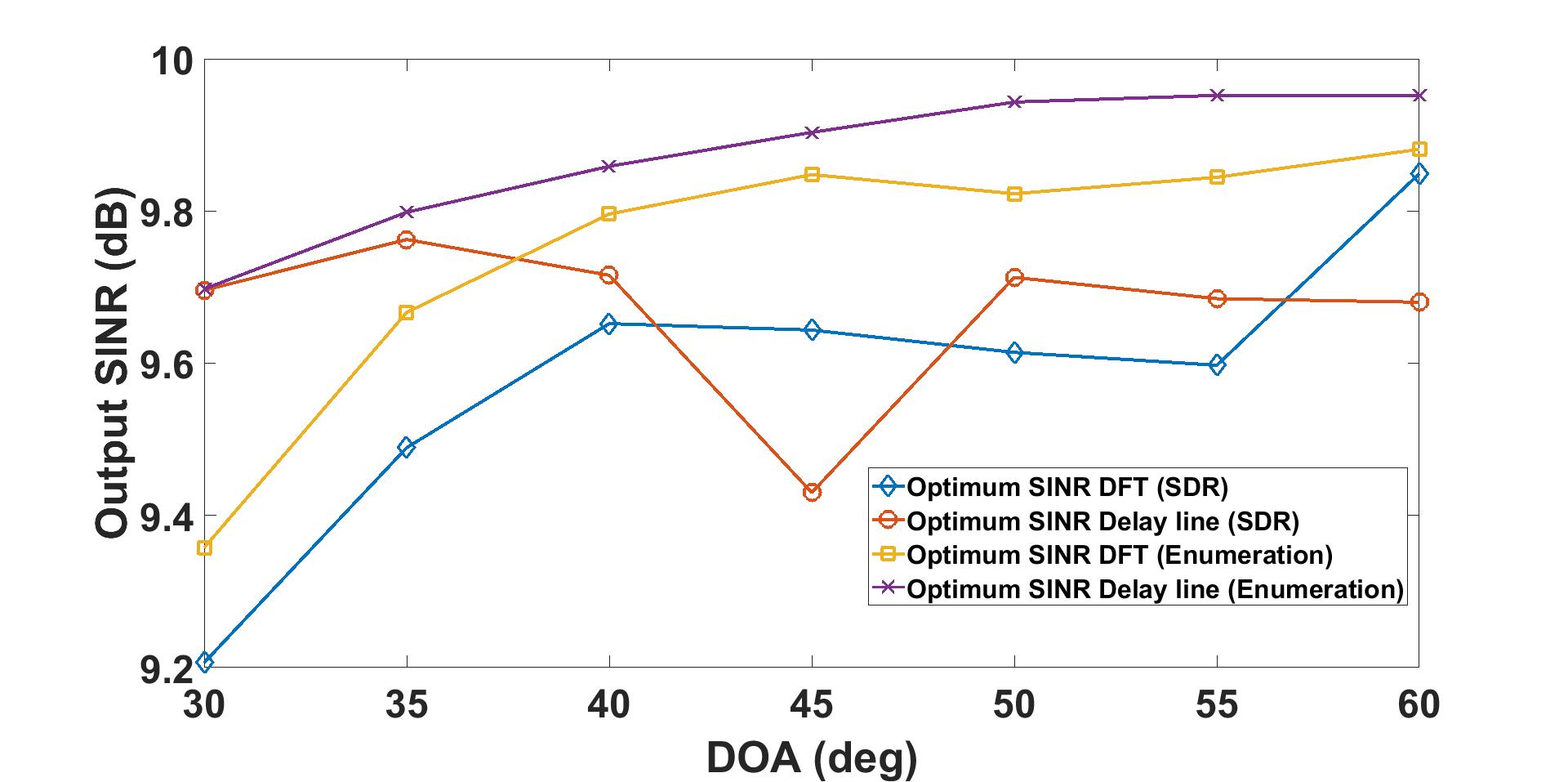}
	\caption{Performance Comparisons of sparse array design for wideband DFT beamformers vs delay line filtering implementation (DOA refers to the angle of source of interest). }
	\label{uni_uni_1_81}
\end{figure}

We  compare the performance of  the DFT and delay line implementations under different operating scenarios. To simulate these scenarios, we shift the DOAs of the above mentioned case in steps of $5^0$. For example, when the desired source is moved from $60^0$ to $55^0$, the corresponding jammers locations are also moved $5^0$ to the left. In this way, we generate seven different operating environments with the source of interest moving from $60^0$ to $30^0$. Figure  (\ref{uni_uni_1_81}) compares the performance of DFT and delay line approach for sparse array  optimization using SDR and by enumeration. The SINR offerings for the delay line approach are higher as compared to the DFT implementation for all operating environments. However, the performance difference is not significant and is associated to the inherent approximation of the orthogonal DFT representation of the signal frequency content. The performance of the SDR algorithm for DFT implementation is comparable to the delay line approach   while being sufficiently closer to the  performance of  sparse array design achieved through enumeration. 

\section{Conclusion}
This paper considered optimum sparse array configuration for maximizing the beamformer output SINR for the case of wideband signal models. It was shown that the sparse array design for the  DFT based wideband beamforming can achieve comparable performance as compared to the delay line approach. The overall dimensionality of the DFT approach is lower for the delay line implementation that renders significant gains in computational complexity. It was found that the weighted mixed  $l_{1-\infty}$-norm squared  group sparsity promoting formulation with principal eigenvector based iterative sparsity control algorithm  is particularly effective in finding the optimum sparse array design with low  computational complexity. We showed the effectiveness of our approach for the frequency spread source operating in wideband jamming environment. The  MaxSINR optimum sparse array design recovered sparse arrays  with comparable performance to the sparse arrays found through  enumeration and showed strong agreement between the two methods.

\balance
\bibliographystyle{IEEEtran}
\bibliography{references}

\end{document}